\documentclass[reqno]{amsart}

\usepackage{amssymb}
\usepackage{amsthm}
\usepackage{longtable}
\usepackage{float}
\usepackage{color}
\usepackage{comment}  
\usepackage{enumitem} 

\newcommand{\nn}{\nonumber}
\usepackage{graphicx}

\newtheorem{theorem}{Theorem}

\numberwithin{equation}{section}
\numberwithin{figure}{section}

\begin{document}

\title[Maps with invariant surfaces]{Rational Maps with Invariant Surfaces}

\author{Nalini Joshi}
\address{School of Mathematics and Statistics F07, The University of Sydney, NSW 2006, Australia}
\email{nalini.joshi@sydney.edu.au}
\thanks{This research was supported by an Australian Laureate Fellowship \# FL 120100094 from the Australian Research Council. }

\author{Claude-Michel Viallet}
\address{ Sorbonne Universit\'e,  Centre National de
  la Recherche Scientifique \\ UMR 7589, LPTHE, 4 Place Jussieu
  \\ F-75252 Paris CEDEX 05, France }
\email{claude.viallet@upmc.fr}
\thanks{}
\bibliographystyle{amsplain}

\date{}

\begin{abstract}
  We provide new examples of integrable rational maps in four
  dimensions with two rational invariants, which have unexpected
  geometric properties, {   as for example orbits confined to
    non algebraic varieties}, and fall outside classes studied by
  earlier authors. We can reconstruct the map from both invariants.
  One of the invariants defines the map unambiguously, while the other
  invariant also defines a new map leading to non trivial fibrations
  of the space of initial conditions.
\end{abstract}

\maketitle

\section{Introduction}\label{section_introduction}
Rational maps in two dimensions with invariant curves form the
starting point for many developments in algebraic geometry and
integrable systems theory. Elliptic curves play a crucial r\^ole in
the development of the
field~\cite{QuRoTh88,QuRoTh89,Sa01,KaMaNoOhYa03,Du10}, see
also~\cite{Mo43,PeSm81}.
{  
Up to now, there has been no general framework to describe integrable maps in
dimension higher than two, but there exists a number of examples,
obtained by various methods. In this paper, we suggest a new starting point and deduce properties of a new class of integrable maps in four dimensions.

In the study of integrable maps, the simplest starting point has been to construct periodic reductions of integrable lattice
equations~\cite{PaNiCa90,KaQu10}, as well as symmetry
reduction~\cite{NiPa91,LeWi06}, since such reductions are
automatically integrable.} A different approach taken in~\cite{PeSu06,PeSu16} was to start from
integrable Hamiltonian systems, and use an appropriate discretisation
to obtain birational maps, together with the adequate symplectic
structure,  ensuring integrability in the sense of Liouville.

Finally, a direct generalisation of the two dimensional case to four
dimensions was given in~\cite{CaSa01}, the idea being to start from
a multiquadratic expression and construct generating involutions
leaving these quantities invariant. Among the resulting maps are
autonomous versions of members of hierarchies of $q$-discrete
Painlev\'e equations~\cite{Ha07}\footnote{For example the autonomous
  version of equation (4.4) in~\cite{Ha07} is the same as equation
  (4.29) in~\cite{CaSa01}.}.

In this paper, we provide new four-dimensional maps with two rational
invariants. The invariants are not multiquadratic, different
coordinates appear with different degrees, and therefore our maps do
not fall into the class described in~\cite{CaSa01}. They arose from
the autonomous limit of additive discrete Painlev\'e
equations~\cite{CrJo99}.  We show that they have vanishing algebraic
entropy~\cite{BeVi99}.{ The vanishing of the entropy will be our test
  of integrability throughout the paper}.
{  
We start by describing our notation. Let $x_n$, for each $n\in\mathbb
Z$, be iterates of a mapping.  We take homogeneous coordinates in four
dimensions in $\mathbb C \mathbb P^4$ to be $[x,y,z,u,t]$, which
stands for $[x_{n}, x_{n-1}, x_{n-2}, x_{n-3}, 1]$ up to a common
factor and define a map $\varphi:\mathbb P^4\to \mathbb P^4$ by the
action
\begin{equation}\label{genaction}
  \varphi: [x_{n}, x_{n-1}, x_{n-2}, x_{n-3}, 1]\mapsto [x_{n+1}, x_{n}, x_{n-1}, x_{n-2}, 1].
  \end{equation}
The two main maps which form the focus of this paper are given by
$\varphi^{\scriptscriptstyle (\rm I) } $ defined by Equation
\eqref{phid4pI}, with action \eqref{eq:d4pI}, and
$\varphi^{\scriptscriptstyle (\rm II) }$ \eqref{phid4pII}, with action
\eqref{eq:d4pII}. Before we state our main result about these maps and
their properties (see Theorem \ref{mainthm}), we give a description of
these results.
}
It is possible to reconstruct the maps from the invariants.  While the
original map can be recovered unambiguously from the lower degree
invariant, the other invariant defines two maps, thus providing us
with an alternate one which we called the {\em shadow} map.  This
phenomenon was described in~\cite{QuCaRo05}, where the case of two
maps in four dimensions was studied, and the additional map was called
a “dual” map.

{   The reason for the existence of this additional map is
  simple.  Maps coming from recurrences are entirely determined by a
  unique equation, and this equation automatically appears as a factor
  in any invariance condition. Depending on the degree structure of
  the invariants, more factors may then arise, each defining a
  different map. For the examples we give here, one invariant (the one
  with the lower degrees) yields the original map, while the other
  invariant (with higher degrees) yields two different maps: original
  and shadow.}

{ For all the cases presented here}, the
shadow map is itself integrable and we show that it leads to a
non-trivial fibration by curves of the 3-fold given by the higher
degree invariant.

Based on these results,  we propose a model of four-dimensional maps
with two rational invariants, and give two more instances of such maps,
one with a structure of invariants similar to the previous ones, and
one presenting novel features.
{  
Our main results are collected in the following theorem.
\begin{theorem}\label{mainthm}
  The maps $\varphi^{\scriptscriptstyle (\rm I) } $ and
  $\varphi^{\scriptscriptstyle (\rm II) } $ are integrable. Each map
  has two explicit homogeneous polynomial invariants given
  respectively by
  \begin{equation}\label{invI}\{t^4\,\Delta_4^{({\rm I})},t^5\,\Delta_5^{({\rm I})} \},
  \end{equation}
  defined in Equations (\ref{deltaI}) and
  \begin{equation}\label{invII}\{t^6\,\Delta_6^{({\rm II})},t^8\,\Delta_8^{({\rm II})} \},\end{equation}
  defined in Equations (\ref{deltaII}). \\
  These maps have the following  properties. 
  \begin{enumerate}[leftmargin=1.2cm,label=(\alph*)]
    \item Each map $\varphi^{\scriptscriptstyle (\rm I) } $ and $\varphi^{\scriptscriptstyle (\rm II) } $ has vanishing algebraic entropy.\\
    \item Both maps and their invariants are unchanged by the involution $\iota : [x,y,z,u,t] \mapsto [u,z,y,x,t].$
    \item The condition that each pair of polynomials \eqref{invI} and \eqref{invII} remains invariant under the iteration \eqref{genaction} gives rise to two new maps $\varphi_s^{\scriptscriptstyle (\rm I)}$ and  $\varphi_s^{\scriptscriptstyle (\rm II)}$, defined respectively by Equations \eqref{phid4pIs} and \eqref{phid4pIIs}. We call these \emph{shadow maps}.
      \item The invariants of each map share a degree pattern in the variables $(x, y, z, u)$. In particular,  $\Delta_4^{({\rm I})}$ and $\Delta_6^{({\rm II})}$ are of degree $(1, 3, 3, 1)$, while $\Delta_5^{({\rm I})}$ and $\Delta_8^{({\rm II})}$ are of degree $(2, 4, 4, 2)$. 
    \end{enumerate}
  \end{theorem}
}
The paper is organised as follows. In Section \ref{dpI}, we describe
the autonomous limit of the second member of the hierarchy of the
discrete first Painlev\'e equation (the equation with initial-value
space $E_6^{(1)}$)~\cite{CrJo99}.  The map is defined from a
recurrence of order 4, and acts on $\mathbb C\mathbb P^4$. We give its
two invariants.  We show that the shadow map is integrable by
calculating its algebraic entropy, giving its three invariants and
deducing a non-trivial elliptically fibered 3-fold from these
results. In Section \ref{dpII}, we provide parallel results for the
second member of the hierarchy of the discrete second Painlev\'e
equation ($D_6^{(1)}$).  Section \ref{structure} describes the
geometry of the invariant surfaces, and gives a construction scheme
for four dimensional maps with two algebraic invariants. Section
\ref{new_one} and \ref{new_two} give two new recurrences, constructed
along the lines of the scheme given in the previous section.  Both are
integrable, but with different characteristics, revealed by the
analysis of the growth of the degrees of their iterates. This
difference is reflected in the nature of the invariants of their
shadow maps: the first shadow map possesses three independent rational
invariants, while the second only has two rational invariants, but
also one non-rational invariant, a situation already encountered
in~\cite{AnMaVi02b}. In Section \ref{an_inflation}, we introduce a
notion of {\em inflation}, which allows us to produce from a
recurrence of order $n$ a new recurrence of order $n+1$. We use this
notion to analyse the model described in Section \ref{new_two}. We
conclude with some directions for further studies.
\bigskip

\section{Autonomous $d_4P^{\scriptscriptstyle (\rm I) }$}
\label{dpI}

In this section, we study the autonomous version of a fourth-order
member of the hierarchy of the discrete first Painlev\'e
equation~\cite{CrJo99}, denoted by $d_4P^{\scriptscriptstyle (\rm I)
}$ (Equation (2.9) of~\cite{CrJo99}). We study the map in $\mathbb C \mathbb
P^4$, by providing invariants, constructing the shadow map and
deducing further properties.

Denoting the iterates by $x_n$, for each $n\in\mathbb Z$, we take
homogeneous coordinates in four dimensions in $\mathbb C \mathbb P^4$
to be $[x,y,z,u,t]$, which stands for $[x_{n}, x_{n-1}, x_{n-2},
  x_{n-3}, 1]$ up to a common factor.  The map then sends
\[
[x_{n}, x_{n-1}, x_{n-2}, x_{n-3}, 1]\mapsto [x_{n+1}, x_{n}, x_{n-1}, x_{n-2}, 1],
\] 
up to common factors. We denote this map by
\begin{eqnarray}
\label{phid4pI}
\varphi^{\scriptscriptstyle (\rm I) } & : & [x,y,z,u,t] \mapsto [x',y',z',u',t'],
\end{eqnarray}
where 
 \begin{equation}\label{eq:d4pI}
   \begin{cases}
 x'  =&  - a\,y \left({x}^{2}+ {y}^{2}+{z}^{2}  +2\,yz+2\,xy
  +xz+zu \right) \\
  &\qquad - b\, t y  \left( y+z+x \right)  -c \, y{t}^{2}+ d\, {t}^{3}, \\
 y' = & a\, y{x}^{2}, \quad z'  = a\, x {y}^{2}, \quad
 u' = a\, xyz, \quad
 t'  = a\, xyt .
\end{cases}
\end{equation}
It can be checked that the map $\varphi^{\scriptscriptstyle (\rm I) }$ has two invariants $\Delta^{\scriptscriptstyle (\rm I) }_4$  and $\Delta^{\scriptscriptstyle (\rm I) }_5$ 
which are:
\begin{subequations}\label{deltaI}
\begin{eqnarray}
\label{delta4}
t^4 \Delta^{\scriptscriptstyle (\rm I) }_4& = & a \; yz \left( -{y}^{2}-2\,yz-xy-{z}^{2}-zu+xu \right)  
  -  b\; tyz \left( z+y \right),  \nn \\
&& -c \; yz{t}^{2}+  d\;{t}^{3} \left( z+y \right) 
\\
t^5 \Delta^{(\scriptscriptstyle I)}_5&  =& 
a\; y z \left( zu+xy+{y}^{2}+2\,yz+{z}^{2} \right)  \left( z+u+y+x \right)
+c\; yz \left( z+u+y +x \right) {t}^{2} \nn
\\ &&
-d \left( zu+xy+{y}^{2}+2\,yz+{z}^{2} \right) {t}^{3}
 +b\; yz \left( y+z+x \right)  
\left( u+y+z \right) t . \label{delta5}
\end{eqnarray}
\end{subequations}
Both invariants are unchanged by the involution
\begin{eqnarray}
  \label{time}
\iota : [ x,y,z,u,t] \mapsto [ u,z,y,x,t].
\end{eqnarray}
The sequence of degrees of the iterates of  $\varphi^{ (\scriptscriptstyle I)}$,
\begin{eqnarray}
  \label{seq}
  \{d_n\}^{ (\scriptscriptstyle I)} = 1,  3, 6, 12, 21, 33, 47, 64, 83, 104, 128, 154, 183, 214, 248, 284, \dots
\end{eqnarray}
is fitted by the rational  generating function
\begin{eqnarray}
\label{genfunc}
g^{ (\scriptscriptstyle I)}(r) = {\frac {{{\it r}}^{10}-{{\it r}}^{9}-{{\it r}}^{6}+2\,{{\it r
}}^{4}+2\,{{\it r}}^{3}+{\it r}+1}{ \left( {\it r}+1 \right) 
 \left( {1-\it r} \right) ^{3}}}.
\end{eqnarray}
The distribution of the poles in (\ref{genfunc}) shows that the
degrees $d_n$ grow polynomially in $n$ with quadratic growth,
implying vanishing of the algebraic entropy.

Remark: the rational nature and the explicit form of the generating
function~(\ref{genfunc}) comes from the fact that the sequence of
degrees~(\ref{seq}) verifies a finite recurrence relation with integer
coefficients. Such a property may be proved along the lines given
in~\cite{Vi15}.

Using the fact that the map is coming from a recurrence, we can
recover the map $\varphi^{(\scriptscriptstyle I)}$ from each
invariant.  In particular, $\Delta^{\scriptscriptstyle (\rm I) }_4( [
  x',x,y,z,t] ) - \Delta^{\scriptscriptstyle (\rm I) }_4( [
  x,y,z,u,t])$ decomposes into two factors, one being trivial ($x-z$)
and the other giving back the map (\ref{phid4pI}). The similar
difference constructed with the invariant $\Delta^{\scriptscriptstyle
  (\rm I)}_5$ also decomposes into two factors, both now giving
nonlinear maps, as observed in~\cite{QuCaRo05}. One is the original map
(\ref{phid4pI}), while the other one turns out to be
\begin{eqnarray}
\label{phid4pIs}
\varphi^{\scriptscriptstyle (\rm I) }_s: [x,y,z,u,t] \mapsto [yz+{z}^{2}+zu-{x}^{2}-xy,{x}^{2},xy,xz,tx],
\end{eqnarray}

The integrability of $\varphi^{\scriptscriptstyle (\rm I) }_s$ can be seen from the evaluation of
its algebraic entropy. The sequence of degrees of its iterates 
\begin{eqnarray*}
 \{d_n\}_s^{ (\scriptscriptstyle I)} = 1, 2, 4, 7, 11, 17, 24, 32, 41, 52, 64, 77, 91, 107, 124, 142  \dots ,
\end{eqnarray*}
 has a generating function given by the rational fraction
\begin{eqnarray}
  g_s^{ (\scriptscriptstyle I)}(r) = 
{\frac {1+{r}^{2}+{r}^{3}+2\,{r}^{5}}{ \left(1+ r \right)  \left(1+ {r}
^{2} \right)  \left( 1-r \right) ^{3}}},
\end{eqnarray}
showing again quadratic growth and vanishing of the  entropy.

The  map $\varphi^{\scriptscriptstyle (\rm I) }_s$   possesses three independent rational invariants:
\begin{subequations}
\begin{eqnarray}
\Sigma^{\scriptscriptstyle (\rm I) }_2 & = &{\frac {zu+xy+{y}^{2}+2\,yz+{z}^{2}}{{t}^{2}}}, \\
\Sigma^{\scriptscriptstyle  (\rm I) }_3 &=&{\frac {yz \left( z+u+y+x \right) }{{t}^{3}}}, \\
\Sigma^{\scriptscriptstyle (\rm I) }_4 & = &{\frac {yz \left( y+z+x \right)  \left( u+y+z \right) }{{t}^{4}}}.
\end{eqnarray}
\end{subequations}
There is  a simple algebraic relation between
$\Delta^{\scriptscriptstyle  (\rm I)}_5$ and the invariants
$\Sigma^{\scriptscriptstyle  (\rm I) }_i$:
\begin{eqnarray}
  \Delta^{\scriptscriptstyle (\rm I)}_5 = a \; \Sigma^{\scriptscriptstyle (\rm I) }_2 \; \Sigma^{\scriptscriptstyle  (\rm I)}_3 + b \; \Sigma^{\scriptscriptstyle (\rm I) }_4 + c \;\Sigma^{\scriptscriptstyle (\rm I) }_3 - d\; 
\Sigma^{\scriptscriptstyle (\rm I) }_2.
\end{eqnarray}
The three invariants $\Sigma^{\scriptscriptstyle (\rm I) }_i$ define a
non-trivial elliptic fibration of the 3-folds of constant
$\Delta_5^{\scriptscriptstyle (\rm I)}$. Indeed the curves in $\mathbb
C \mathbb P_4$ defined by $\Sigma_i^{ \scriptscriptstyle(\rm I)} =
constant, i =2,3,4$ have an infinite group of automorphisms
provided by $\varphi_s$ itself, and are consequently of genus $g\leq
1$. 

Furthermore, the compositions $\tau = \iota \cdot
\varphi^{\scriptscriptstyle (\rm I)}$ and $\tau_s = \iota \cdot
\varphi^{\scriptscriptstyle (\rm I)}_s$ define further involutions, which
moreover commute, i.e.  $\tau \cdot \tau_s = \tau_s \cdot \tau$.

Remark: Note that any functional combination of
$\Delta^{(\scriptscriptstyle I)}_5 $ and $\Delta^{\scriptscriptstyle
  (\rm I)}_4$ will also be an invariant of
$\varphi^{(\scriptscriptstyle I)}$. Different choices will lead to
different shadow maps (see~\cite{QuCaRo05}). The choices we made above
were based on two requirements: firstly, to produce invariants of
minimal degree, and secondly to define the simplest possible shadow
map.

\section{Autonomous $d_4P^{\scriptscriptstyle (\rm II) }$}
\label{dpII}
In this section, we study the autonomous version of a fourth-order member of the hierarchy of the discrete second Painlev\'e equation~\cite{CrJo99}, denoted by $d_4P^{\scriptscriptstyle (\rm II) }$. (The latter is Equation (3.7) of~\cite{CrJo99}.) 

In $\mathbb C \mathbb P^4$, this equation gives rise to the map  
\begin{eqnarray}
\label{phid4pII}
\varphi^{\scriptscriptstyle  (\rm II) } & : & [x,y,z,u,t] \mapsto [x',y',z',u',t'],
\end{eqnarray}
 with 
\begin{eqnarray}\label{eq:d4pII}
\begin{cases}
 x'  =&  d \;{t}^{5} - a \; \left( t-y \right)  \left( t+y \right)  \left( u{t}^{2}-y{z}^{2}-u{z}^{2}-2\,yxz-{x}^{2}y \right)  -c\; y {t}^{4}  \\
 & - b \; {t}^{2} \left( t-y \right)  \left( t+y \right)  \left( z+x \right), 
\\
 y' =& ax \left( {t}^{2}-{y}^{2} \right)  \left( {t}^{2}-{x}^{2} \right) , 
\quad z'  = ay \left( {t}^{2}-{y}^{2} \right)  \left( {t}^{2}-{x}^{2} \right) , 
\\ 
 u' =& az \left( {t}^{2}-{y}^{2} \right)  \left( {t}^{2}-{x}^{2} \right) , 
\quad
  t'  = at \left( {t}^{2}-{y}^{2} \right)  \left( {t}^{2}-{x}^{2} \right).
\end{cases}
\end{eqnarray}
The invariants $\Delta^{\scriptscriptstyle (\rm II) }_6$  and $\Delta^{\scriptscriptstyle (\rm II) }_8$ of $\varphi^{\scriptscriptstyle (\rm II) }$ are given by
\begin{subequations}\label{deltaII}
\begin{eqnarray} \nn
t^6 \Delta^{\scriptscriptstyle (\rm II) }_6& = &  a\; \left( t-z \right)  \left( t+z \right)  \left( t-y \right)  \left( t+y \right)  \left( ux-uz-xy-yz \right)  
\\ &&  
-b\; {t}^{2} \left( {z}^{2}{t}^{2}+
{t}^{2}{y}^{2}-{z}^{2}{y}^{2} \right)  -c\; {t}^{4}yz
+ d \; {t}^{5} \left( z+y \right),
\\
t^8 \Delta^{\scriptscriptstyle (\rm II) }_8&  =& a \; \bigl( 
 \left( {u}^{2}+{z}^{2}+{y}^{2}+{x}^{2} \right) {t}^{6} -{z}^{2}{y}^{2} \left( uz+xy+yz \right) ^{2} \nn
\\ && 
- \left( 2\,yu{z}^{2}+2\,uzxy+{x}^{2}{z}^{2}+2\,xz{y}^{2}+2\,{x}^{2}{y
}^{2}+{u}^{2}{y}^{2}+2\,{z}^{2}{y}^{2}+2\,{u}^{2}{z}^{2} \right) {t}^{4} \nn
\\ &&
+ ( 2\,{x}^{2}{y}^{2}{z}^{2}+2\,u{y}^{3}{z}^{2}+2\,x{y}^{2}{z}^{3}
+2\,yu{z}^{4}+{z}^{2}{y}^{4}+{y}^{2}{z}^{4}+2\,{u}^{2}{y}^{2}{z}^{2}
+{x}^{2}{y}^{4} \nn
\\ &&
+2\,ux{y}^{3}z+2\,uxy{z}^{3}+2\,xz{y}^{4}
+{z}^{4}{u}^{2} ) {t}^{2}\bigr)\nn
\\ &&
+ b \; {t}^{2} \left( t-z \right)  \left( t+z \right)  \left( t-y \right)  
\left( t+y \right)  \left( z+x \right)  \left( u+y \right) \nn
\\ &&
+ c\; {t}^{4} \left( xz{t}^{2}-{z}^{2}{y}^{2}+yu{t}^{2}-yu{z}^{2}
-xz{y}^{2} \right)  \nn
\\ &&
- d\; {t}^{5} \left( x{t}^{2}+z{t}^{2}-z{y}^{2}
-x{y}^{2}-u{z}^{2}+u{t}^{2}-y{z}^{2}+y{t}^{2} \right).
\end{eqnarray}
\end{subequations}
Both invariants are again unchanged by the involution
\begin{eqnarray}
\iota : [ x,y,z,u,t] \mapsto [ u,z,y,x,t].
\end{eqnarray}

In this case, the shadow map $ \varphi^{\scriptscriptstyle (\rm II) }_s$, defined as above, is
\begin{eqnarray}\label{phid4pIIs}
  [x,y,z,u,t] \stackrel {  \varphi^{\scriptscriptstyle (\rm II) }_s}  { \mapsto}
 &[{x}^{2}y-y{z}^{2}-u{z}^{2}+u{t}^{2},  x \left( {t}^{2}-{x}^{2} \right) 
  ,\nn \\
  &\qquad y \left( {t}^{2}-{x}^{2} \right) , 
z \left( {t}^{2}-{x}^{2} \right) ,t
  \left( {t}^{2}-{x}^{2} \right) ]
\end{eqnarray}
and it turns out to also have vanishing algebraic entropy.

We find  that $ {  \varphi^{\scriptscriptstyle (\rm II) }_s}$ has three independent rational invariants
\begin{eqnarray*}
\Sigma^{\scriptscriptstyle (\rm II) }_3 & = &  {\frac {x{y}^{2}+z{y}^{2}+u{z}^{2}
+y{z}^{2}+ \left( -z-y-x-u \right) {t}^{2}}{{t}^{3}}}, \\
\Sigma^{\scriptscriptstyle (\rm II) }_4 & = & {\frac { \left( zx+uy \right) {t}^{2}
-yz \left( uz+xy+yz \right) }{{t}^{4}}},
\\
\Sigma^{\scriptscriptstyle  (\rm II)}_6 & = &{\frac { \left( t-z \right)  
\left( t+z \right)  \left( t-y \right) 
 \left( t+y \right)  \left( z+x \right)  \left( u+y \right) }{{t}^{6}}}.
\end{eqnarray*}
The invariant $\Delta^{\scriptscriptstyle (\rm II) }_8$ has a  simple algebraic relation to the $\Sigma^{\scriptscriptstyle (\rm II)}_j$: 
\begin{eqnarray*}
\Delta^{\scriptscriptstyle (\rm II) }_8 = a \;( {\Sigma^{\scriptscriptstyle (\rm II)}_3}^2 -2 {\Sigma}_4^{\scriptscriptstyle  (\rm II)} -  {\Sigma_4^{\scriptscriptstyle (\rm II)}}^2 - 2 \Sigma^{\scriptscriptstyle (\rm II) }_6) + b \; \Sigma^{\scriptscriptstyle (\rm II) }_6 + c\;  \Sigma^{\scriptscriptstyle (\rm II)}_4 + d\; \Sigma^{\scriptscriptstyle (\rm II) }_3 .
\end{eqnarray*}
The situation is very similar to the previous section.

\section{On the structure of the invariants}
\label{structure}
In this section, we study the structure of the rational invariants
given in the previous two sections. Our starting point is the
distribution of degrees shared by the pair of invariants arising from
autonomous $d_4P^{\scriptscriptstyle(\rm I)}$ and that arising from
$d_4P^{\scriptscriptstyle(\rm II)}$. We show that their properties
give rise to ruled 3-folds and elliptic 3-folds.  We extend these
properties to new rational invariants with similar structures.

The invariants in the previous sections are ratios of homogeneous
polynomials. These polynomials have well defined degrees with respect
to the homogeneous coordinates $[x,y,z,u,t]$. For both models we gave
the simplest form of the invariants, where the denominators are just
powers of $t$.

While the invariants of each model have different total degree ($4$
and $5$ for $d_4P^{\scriptscriptstyle(\rm I)}$, $6$ and $8$ for
$d_4P^{\scriptscriptstyle(\rm II)}$), they share the same structure:
the distribution of degrees with respect to each variable is similar
for the numerators of $\Delta_4^{\scriptscriptstyle(\rm I)}$ and
$\Delta_6^{\scriptscriptstyle(\rm II)}$ (respectively for
$\Delta_5^{\scriptscriptstyle(\rm I)}$ and
$\Delta_8^{\scriptscriptstyle(\rm II)}$). See Table \ref{tab1}.

\begin{table}[H]
\begin{center}
\label{degrees}
\begin{tabular}{|c| cccc| c |c|} 
\hline
 &  $x$ & $y$ & $z$ & $u$ & $t$ & Total degree \\
\hline
 $\Delta_4^{\scriptscriptstyle(\rm I)}$& 1 & 3 & 3 & 1 & 3 & 4 \\
 $\Delta_6^{\scriptscriptstyle(\rm II)}$& 1 & 3 & 3 & 1 & 5 & 6 \\
\hline
 $\Delta_5^{\scriptscriptstyle(\rm I)}$& 2 & 4 & 4 & 2 & 3 & 5 \\
 $\Delta_8^{\scriptscriptstyle(\rm II)}$& 2 & 4 & 4 & 2 & 7 & 8 \\
\hline
\end{tabular}
\end{center}
\caption{The distribution of degrees of invariants.}
\label{tab1}
\end{table}

Notice that $\Delta_4^{\scriptscriptstyle(\rm I)}$ and $\Delta_6^{\scriptscriptstyle(\rm II)}$ are linear in $x$ and $u$. Therefore, the varieties $\Delta_4^{\scriptscriptstyle(\rm I)} = constant$ (resp.
$\Delta_6^{\scriptscriptstyle(\rm II)} = constant$) have intersections with the $y$ and $z$-coordinate hyperplanes that are straight lines. Therefore, these varieties are ruled 3-folds. 

On the other hand, notice that $\Delta_5^{\scriptscriptstyle(\rm I)}$ and $\Delta_8^{\scriptscriptstyle(\rm II)}$ are quadratic in $x$ and $u$. 
Therefore, there is an elliptic fibration of
the 3-dimensional varieties $\Delta_5^{\scriptscriptstyle(\rm I)} = constant$ (resp.
$\Delta_8^{\scriptscriptstyle(\rm II)} = constant$).

For both previous cases, the shadow maps provide us with a non trivial
elliptic fibration of the varieties $\Delta_5^{\scriptscriptstyle(\rm
  I)} = constant$ (resp.  $\Delta_8^{\scriptscriptstyle(\rm II)} =
constant$). Indeed their orbits are confined to algebraic curves
defined by the invariants $\Sigma_j^{\scriptscriptstyle(\rm I)}$
(resp.  $\Sigma_k^{\scriptscriptstyle(\rm II)}$). We know these curves
are elliptic (or accidentally rational) curves since they possess an
infinite group of automorphisms: the iterates of the shadow map.

To construct new recurrences of the same type, we will proceed as follows.
\begin{enumerate}
\item[1.] Choose two polynomials $\Pi_l$ and $\Pi_h$ with respective
  degrees $1,3,3,1$ and $2,4,4,2$ in $x,y,z,u$ as in
  Table~\ref{tab1}, and total  degrees   $d_l$ and $d_h$.
\item[2.] Impose the condition that both polynomials are invariant
    by the involution $\iota$ (Equation~\eqref{time}), and define $\Delta_l =
    \Pi_l/t^{d_l}$ and $\Delta_h=\Pi_h/t^{d_h}$.
\item[3.] Assume that the conservation condition of $\Delta_l $  factors as
\begin{eqnarray}
\label{main}
\Pi_l( x',x,y,z,t) - \Pi_l(x,y,z,u,t) = ( x-z ) \; Q( x', x, y, z, u, t).
\end{eqnarray}
We thus get from Equation \eqref{main} a recurrence relation, and a
birational map
\begin{eqnarray}
\varphi: [x,y,z,u,t] \mapsto  [x',y',z',u',t'].
\end{eqnarray}
\item[4.]  Impose the condition that the higher degree polynomial
  $\Pi_h$ verifies:
\begin{eqnarray}
  \Pi_h( x',x,y,z,t) - \Pi_h(x,y,z,u,t) = Q( x', x, y, z, u, t) \; S( x',
  x, y, z, u, t).
\end{eqnarray}
The previous condition ensures the invariance of
$\Delta_h$ under $\varphi$ and defines the shadow map
$\varphi_s$ by solving $S( x', x, y, z, u, t)=0$.
\end{enumerate}

Remark: Denoting by $\delta(R,*)$ the discriminant of $R$ with respect to variable $*$,
$\Pi_h$ verifies the necessary condition
\begin{eqnarray}
\delta( \delta ( \Pi_h, x) , u ) =0.
\end{eqnarray}

In what follows, we study examples of pairs of polynomials $(\Pi_l, \Pi_h)$
solving these conditions, both with the minimal total degrees $d_l=3$ and
$d_h=4$.

\section{Another elliptic fibration}
\label{new_one}
In this section, we provide an example of a pair of polynomials $(\Pi_l, \Pi_h)$ satisfying the conditions given in Section \ref{structure}.
Define two polynomials
\begin{eqnarray*}
  \Pi_l^{(1)} &=& a\; {t}^{2} \left( z+y \right) + b\; t \left( {z}^{2}
+{y}^{2} \right)   + c \left( z+y \right)  \left( xu-zu-xy-yz \right) \\
&&-t \left( xu-zu-xy-    3\,yz \right),
  \\
 \Pi_h^{(1)} &=&  
a\;c\; t^2 \left( uy+zu+xy+xz+2\,yz \right) 
-a\;{t}^{3} \left( z+u+y+x \right) \\
&&+b\;c\;t\; \left( z+y \right)  \left( x+z \right)  \left( u+y \right) 
- b\;{t}^{2} \left( x+z \right)  \left( u+y \right)\\
&&  
- c^2 \Bigl( {u}^{2}{y}^{2}+2\,{u}^{2}yz+{u}^{2}{z}^{2}+2\,u{y}^{2}z
+2\,y{z}^{2}u+{x}^{2}{y}^{2}+2\,{x}^{2}yz\\
&&\qquad +{x}^{2}{z}^{2}
+2\,x{y}^{2}z+2\,xy{z}^{2}+2\,{y}^{2}{z}^{2} \Bigr) \\
&&
+2\,c\;t \Bigl( {u}^{2}y+{u}^{2}z+u{y}^{2}+2\,zuy+y{x}^{2}+{x}^{2}z+2\,xyz
+x{z}^{2}\\
&&\qquad +{y}^{2}z+y{z}^{2} \Bigr)\\
&&-{t}^{2} \left( {u}^{2}+2\,uy+{x}^{2}+2\,xz+{y}^{2}+{z}^{2} \right).
\end{eqnarray*}
The corresponding map $\varphi^{(1)} $ and  shadow map
$\varphi_s^{(1)} $ are given respectively  by
\begin{eqnarray}
\varphi^{(1)}(x, y, z, u, t):\begin{cases}
x' =& a\;{t}^{2}+b\; t \left( x+z \right) 
-c \left( xy+yz+uy+zu \right) +t \left( 2\,y+u \right) ,\\
y' =& c\;x \left( x+y \right) -x\;t,\qquad
z' = c\; y \left( x+y \right) -y\; t, \\
 u' =& c\; z \left( x+y \right) -z\; t, \qquad
t' = c\; \left( x+y \right) t-{t}^{2},
\end{cases}
\end{eqnarray}
and 
\begin{eqnarray} 
\varphi_s^{(1)}(x, y, z, u, t):\begin{cases}
 x' =& -u\; t -  c\; \left( xy-yz-uy-zu \right)  ,\\
y' =& c\;x \left( x+y \right)  - x\;t ,\qquad
z' =  c\;y \left( x+y \right) - y\; t, \\
u' =& c\; z \left(    x+y \right) - z\; t,\qquad
t =  c\; t \left( x+y \right) - {t}^{2}
\end{cases}
\end{eqnarray}
The sequence of degrees of the iterates are:
\begin{eqnarray*}
\{d_n\}^{(1)}& =& 1, 2, 4, 8, 14, 22, 32, 44, 57, 72, 88, 106, 126, 148, 172, 198 , \dots \\
\{d_n\}_s^{(1)} &=& 1,2,4,8,14,21,30,40,52,66,81,98,116,136, 158, 181, \dots
\end{eqnarray*}
fitted by the rational generating functions
\begin{eqnarray*}
g^{(1)}(r) = {\frac {{r}^{11}-{r}^{10}+{r}^{9}-{r}^{8}+{r}^{3}+{r}^{2}-r+1}{
 \left( 1-r \right) ^{3}}} \quad {\mbox {and}}
\end{eqnarray*}
\begin{eqnarray*}
g_s^{(1)}(r)={\frac {2\,{r}^{6}+2\,{r}^{4}+2\,{r}^{3}+{r}^{2}+1}{ \left( {r}^{4}+{
r}^{3}+{r}^{2}+r+1 \right)  \left( 1-r \right) ^{3}}},
\end{eqnarray*}
showing quadratic growth and, therefore, integrability.
Again, $\varphi_s^{(1)}$ has three independent rational invariants
$\Sigma_i^{(1)}$
\begin{eqnarray*}
t^2 \Sigma_2^{(1)}& =& c\; \left( uy+zu+xy+xz+2\,yz \right) 
-t \left( z+u+y+x \right) ,
\\
t^3 \Sigma_3^{(1)} &=& \left( x+z \right)  \left( u+y \right)  \left( c \left( z+y
 \right) -t \right) ,
\\
t^4 \Sigma_4^{(1)} &=&  
c\;t \left( u{z}^{2}+x{y}^{2}+{y}^{2}z+y{z}^{2} \right)
+ c^2\left( xy+xz+yz \right)  \left( uy+zu+yz \right)\\
&&\quad
-{t}^{2} \left( x+z \right)  \left( u+y \right) ,
\end{eqnarray*}
and $\Delta_h^{(1)}=\Pi_h^{(1]}/t^h$ can be expressed in terms of the $\Sigma_i^{(1)}$:
\begin{eqnarray}
\Delta_h^{(1)} = a \;\Sigma_2^{(1)} - (\Sigma_2^{(1)})^2 +(b-4) \Sigma_3^{(1)} 
+2 \;\Sigma_4^{(1)}.
\end{eqnarray}
The orbits of  $\varphi_s^{(1)}$ are confined to elliptic curves which
provide us with an elliptic fibration of the variety $\Delta_h^{(1)} =
constant$. The situation is very similar to the one encountered in
sections \ref{dpI} and \ref{dpII}.

\section{Beyond elliptic fibrations}
\label{new_two}
In this section, we show that not all shadow maps arising from polynomials of the type defined in Section \ref{structure} result in elliptic fibrations.
Consider the two polynomials:
\begin{eqnarray*}
  \Pi_l^{(2)} &=& a\; t \left( z+y \right) ^{2}
+b \left( z+y \right) {t}^{2}
- c\;t \left(x -z \right)  \left(u -y \right) \\
&&
+ \left( z+y \right)  \left( xu-zu-xy-2\,{y}^{2}-3\,yz-2\,{z}^{2} \right),
  \\
 \Pi_h^{(2)} &=& 
a\;c\; t^2 \left( xu+2\,uy+zu+xy+2\,xz+yz \right)
-a\;t \left( z+y \right)  \left( 2\,y+x+z \right)  \left( u+y+2\,z \right) 
\\
 & &
 +b\; c   \left( z+u+y+x \right) {t}^{3}
-b\; t^2 \left( z+y \right)  \left( x+2\,z+u+2\,y \right) \\ &&
+c^2\;t^2 \left( {u}^{2}-2\,uy+{x}^{2}-2\,xz+{y}^{2}+{z}^{2} \right) 
+ \left( z+y \right) ^{2} \left( x+2\,z+u+2\,y \right) ^{2}
\\ &&
-2\,c\;t  \left( z+y \right)  \left( {u}^{2}+xu+uy+2\,zu+{x}^{2}+2\,xy+xz-yz \right) .
\end{eqnarray*}
The corresponding map $\varphi^{(2)} $ and shadow map $\varphi_s^{(2)}
$ are given by
\begin{eqnarray} \label{phi10}
\varphi^{(2)}(x, y, z, u, t):\begin{cases}
x' =& -a \;t \left( 2\,y+x+z \right) -b\;{t}^{2}- c \;t \left(u -2\,y \right) 
+4\,{y}^{2}\\
&\quad +2\,({x}^{2}+xz+z^2) +5\,y(x+z)+u( y +z),\\
y'=& x \left( c\, t-x-y \right) ,
\qquad
z' = y \left(c\, t-x-y \right) ,
\\
 u' =& z \left( c\, t-x-y \right) ,
\qquad
t' = t \left( c\, t-x-y \right) ,
\end{cases}
\end{eqnarray}
and 
\begin{eqnarray} \label{phialt10}
\varphi_s^{(2)}(x, y, z, u, t):\begin{cases}
 x' =&  c\;u\,t+2\,({x}^{2}-z^2) +3\,y\,(x-z)-u(y+z),\\
y' =& x \left(c\,  t-x-y \right) ,\qquad
z' = y \left( c\, t-x-y \right) ,\\
u' =& z \left( c\, t-x-y \right) ,\qquad
t' =  t \left( c\, t-x-y \right).
\end{cases}
\end{eqnarray}
The sequence of degrees of the iterates is now
\begin{eqnarray*}
  \{d_n\}^{(2)} = 1,2,4,8,13,21,31,45,61,82,106,136,169,209,253, 305,361, 426
  , 496, 576, 661,    \dots 
\end{eqnarray*}
and
\begin{eqnarray*}
  \{d_n\}_s^{(2)} = 1 , 2, 4, 7, 12, 19, 28, 40, 55, 73, 95, 121, 151, 
  186, 226, 271, 322, 379, 442, 512, 589,  \dots
\end{eqnarray*}
\begin{figure}[H]
\includegraphics[scale=0.5]{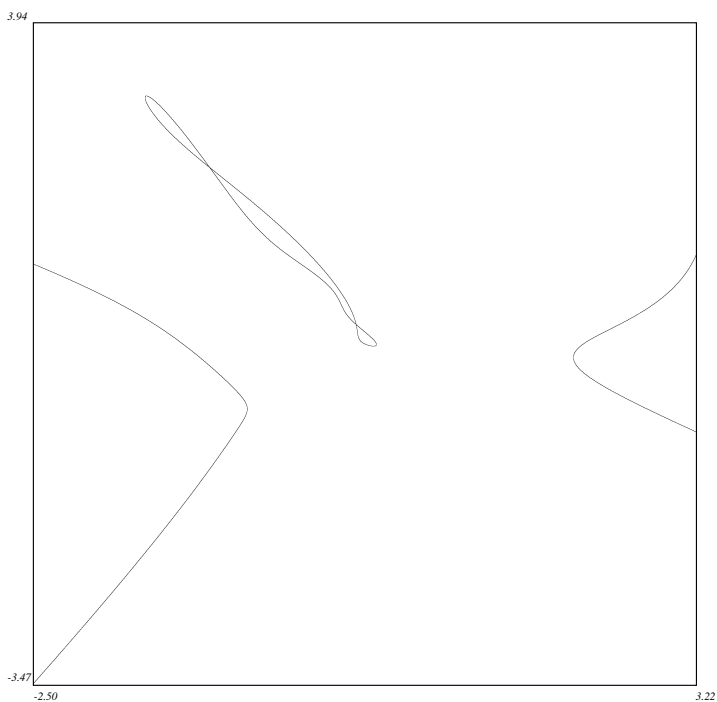}
\caption{An orbit of  $\varphi_s^{(2)}$}
\label{fig2}
\end{figure}
These two sequence are fitted by the generating functions
\begin{eqnarray}
g^{(2)}(r) =  {\frac { 1  +2\,r^3 - r^4 + r^5 + r^6 - r^7 }{ \left( {r}^{2}
+1 \right)  \left( r+1 \right) ^{2} \left( r-1 \right) ^{4}}},
\end{eqnarray}
 and
\begin{eqnarray}
g^{(2)}_s(r) = {\frac {1 - r + r^2 - r^3 + 2\,r^4 -r^5}{ \left( {r}^{2}+r+1
 \right)  \left( r-1 \right) ^{4}}}.
\end{eqnarray}

The fact that the poles of $g(r)$ and $g_s(r)$ at unity are of order 4
means that the two above sequences have {\em cubic growth}.

The  map  $\varphi_s^{(2)}$  has {\em only two} independent rational invariants
$\Sigma_2^{(2)}$ and $\Sigma_3^{(2)}$:
\begin{eqnarray*}
t^2 \Sigma_2^{(2)}& =&  c \left( z+u+y+x \right) t- \left( z+y \right)  \left( x+2\,z +u+2\,y \right),
\\
t^3 \Sigma_3^{(2)} &=& 2\,{c}^{2} t^2 \left( z+u+y+x \right) 
+c\; t \left( xu-zu-xy-4 \,{y}^{2}-7\,yz-4\,{z}^{2} \right) , \\
&& 
-  \left( z+y \right)  \left( 2\,y+x+z \right)  
\left( u+y+2\,z \right) 
\end{eqnarray*}
and $\Delta_h^{(2)}=\Pi_h^{(2]}/t^h$ can be expressed in terms of the
  $\Sigma_j^{(2)}$:
\begin{eqnarray}
\Delta_h^{(2)} = (\Sigma_2^{(2)})^2 + ( 2ac-4c^2-b) \; \Sigma_2^{(2)} 
- (a-2c)\; \Sigma_3^{(2)}.
\end{eqnarray}
At this point drawing a typical orbit of $\varphi_s^{(2)}$ is
extremely useful. See Figure \ref{fig2} for a projection of an orbit
on a 2-dimensional plane.  The picture of a generic orbit shows that
there exists an additional invariant of the shadow map.  This third
invariant cannot be rational. Indeed, if it was rational the orbits
would be confined to elliptic curves {\em and the growth of the degree
  of the iterates would be quadratic, not cubic}~\cite{Be99,Gi80}.

The orbits of the shadow map are not confined to elliptic curves, but
to non algebraic curves drawn on the two dimensional  algebraic variety
$\Sigma_2^{(2)} = constant$, $ \Sigma_3^{(2)} = constant$.

Remark: Maps with a cubic degree growth such as the ones described in
this section {\em cannot be obtained by a reduction from any of the
  known integrable quad equation, since these quad equations all lead
  to a quadratic degree growth}.

\section{An inflation process}
\label{an_inflation}
The two maps of the previous section are related to algebraically
integrable models, by a simple inflation process, defined as follows.

Given an integrable recurrence of order $N$ defined on a variable
$x_n$, leading to a birational map $\varphi$ on $\mathbb C \mathbb
P_N$, one may ``inflate'' the recurrence to order $N+1$ on a new
variable $y_n$ related to $x_n$ by
\begin{eqnarray}
  \label{inflation}
  x_n =\frac{ \alpha_1\; y_n\, y_{n-1} + \alpha_2\; y_n + \alpha_3
    \;y_{n-1} + \alpha_4}{\alpha_5 \; y_n \, y_{n-1} + \alpha_6\; y_n
    + \alpha_7\; y_{n-1} + \alpha_8}
\end{eqnarray}
with  $\alpha_i,{ i=1 \dots 8}$ arbitrary constants.

Remark 1: With specific choices of the parameters $\alpha_i$, Equation
(\ref{inflation}) is known to appear in various transformations, among
which are the definition of potential forms, the so-called discrete
Cole-Hopf transformation~\cite{LeRaBr82}, the discrete Miura
transformation (see for example Equation (1) in~\cite{JoRaGr98}) {   and
the "Gambier coupling" (Equation (3.1) in~\cite{LaGrRa98})}.  Such
transformations act non-trivially, as they are not just coordinate
transformations.

Remark 2: Going from the order $N$ recurrence on $x_n$ to the order
$N+1$ recurrence on $y_n$ can be undone by a ``deflation'' transform,
going from the recurrence on $y_n$ to the one on $x_n$. While
inflation is always possible, deflation cannot be done on arbitrary
recurrences.

The recurrence on $y_n$ defines a birational map $ \varphi^+$ on
$\mathbb C \mathbb P_{N+1}$. Even if the entropy is preserved by
inflation, the sequence of degrees of the iterates is not.  In the
integrable case, where the degrees of the iterates of $\varphi$ grow
polynomially, the inflated map $\varphi^+$ will then still have
vanishing entropy, but {\em possibly with a different polynomial
  growth}. This is what happens for the two maps of the previous
section, with the simple redefinition
\begin{eqnarray}
  \label{somme}
  x_n = y_n + y_{n-1}.
\end{eqnarray}

The map (\ref{phi10}) in $\mathbb C \mathbb P_4$ is obtained by
(\ref{somme}) from the following map in $\mathbb C \mathbb P_3$:
\begin{eqnarray}
&&  [x,y,z,t]\mapsto [x',y',z',t'] \qquad \mbox{  with} \nonumber \\
&&  \begin{cases}
    x' = & - a\, t \left( x+y \right) -b\, c \, {t}^{2}+t \left( x+y-z \right) 
    +{x}^{2}+2\,xy+{y}^{2}+yz \\
    y' = & x\; \left( c\, t-x \right), \qquad
z' = y \; \left( c\, t-x \right), \qquad
t'=  t \; \left( c\, t-x \right) .
 \end{cases} 
  \end{eqnarray}
The latter has {\em quadratic growth} and two rational invariants $\Gamma_1$
and $\Gamma_2$
\begin{eqnarray*}
  t^3\; \Gamma_1& =& -a\, t{y}^{2}-b{t}^{2}y- c\;\left( y-z \right)
  \left( x-y \right) t+y \left( xy-xz+{y}^{2}+yz \right) ,
\\
t^4 \; \Gamma_2 &=&
-{a}^{2}{t}^{2}{y}^{2}-a\, b\, {t}^{3}y+2\,ac{t}^{2}xz-2\,atxyz
+ b c\, {t}^{3} \left( x+  z \right)
  \\
\hskip -1truecm  &&
  +c^2 {t}^{2} \left( {x}^{2}-2\,xy+2
  \,{y}^{2}-2\,yz+{z}^{2} \right)
  -2 c\,yt \left( {x}^{2}+xz-{y}^{2}+{z} ^{2} \right)
  \\
  &&
  -b \,{t}^{2}y \left( x+y+z \right)
  +{y}^{2} \left( x+y+z \right) ^{2}  .
\end{eqnarray*}
One may further reduce the order, eliminating $z$ by specifying the
value of the invariant $\Gamma_1 = k_1$. One obtains a birational map of
$\mathbb C \mathbb P_2$ $[x,y,t]\mapsto [x',y',t']$:
\begin{eqnarray}  \label{2Dphi10}
  \begin{cases}
     x' = & - a\, t{x}^{2}- b \,x {t}^{2}+ c \,x t \left( x-y \right) -k_1\,{t}^{3}+{x}^{3}+{x}^{2}y
\\
    y' = & x\, \left(x -y \right)  \left( c\, t-x \right), 
\qquad
    t' =  t \, \left( x-y \right)  \left( c\, t-x \right). 
  \end{cases}
\end{eqnarray}
This map possesses  an invariant $I= J/ t^4( x-y)^2$   with 
\begin{eqnarray*}
  J &=& {a}^{2}{t}^{2}{x}^{2}{y}^{2}+a\, b \left( x+y \right)x y{t}^{3}
+ 2\,a t \left( k_1{t}^{3}{x}^{2}-k_1{t}^{3}xy+k_1{t}^{3}{y
}^{2}-{x}^{3}{y}^{2}-{x}^{2}{y}^{3} \right)
\\
&& +{b}^{2}{t}^{4}xy
+ b c \left( x+y \right)  \left(x -y \right) ^{2}{t}^{3}
+b\, {t}^{2} \left( x+y \right)  \left( k_1{t}^{3}-{x}^{2}y-x{y}^{2} \right)
+ c^2 \left( x-y \right) ^{4}{t}^{   2}
 \\
 &&
 -2 c\, t \,xy \left( x+y \right)  \left(x -y \right) ^{2}+ \left( k_1{t}
^{3}-{x}^{2}y-x{y}^{2} \right) ^{2}
 +2 a \, c \,xy \left(x - y \right) ^{2}{t}^{2}.
\end{eqnarray*}
It is interesting to notice that defining a recurrence in $\mathbb C \mathbb P_2$ from
the invariant $I$ by imposing $I(x',x,t)=I(x,y,t)$ gives
(\ref{2Dphi10}) as well as non rational maps, due to its degree
distribution. This indicates that it may be of a non QRT
type~\cite{Du10}.

The shadow map (\ref{phialt10}) may be understood in a similar way.
The reduction to three dimensions obtained by eliminating u through
the invariance condition $\Sigma_2^{(2)}=k_2$ yields the map in
$\mathbb C \mathbb P_3$ $[x,y,z,t] \mapsto [x',y',z',t']$
\begin{eqnarray}
  \label{3Dalt10}
  \begin{cases}
  &   x'= -c \left( x+y+z \right) t-{\it k_2}\,{t}^{2}+2\,(x+y)^{2} +(x +y)z, \\
    &
    y'=   x \left( c\,t-x-y \right) ,\; z'=y \left( c\,t-x-y \right) ,\;
    t'= t \left( c\,t-x-y \right)
    \end{cases}
\end{eqnarray}
for which the sequence of degrees has cubic growth. One gets
(\ref{3Dalt10}) as the result of the inflation process (\ref{somme})
applied to the known recurrence
\begin{eqnarray}
  \label{dPxx}
    x_{n+1} + x_{n-1} =  \frac{k_2 - x_n^2}{c - x_n}.
  \end{eqnarray}
  In fact, taking $x_n=c-w_n$,  we find
  \[
  w_{n+1}+w_n+w_{n-1}=\frac{\alpha}{w_n}+2 w_n
  \]
  with $\alpha=c^2-k_2$, which is an autonomous version of the discrete
  first Painlev\'e equation $dP_I$ \cite{FoItKi91}.
  
  Remark: (\ref{2Dphi10}), (\ref{3Dalt10}) and (\ref{dPxx}) can be
  made non-autonomous by varying $k_1$ and $k_2$ respectively. This
  turns (\ref{dPxx}) into the first discrete Painlev\'e equation
  $dP_I$.

\section{Conclusion}\label{section:conclusion}

By examining the autonomous limits of the first members of hierarchies
of discrete Painlev\'e equations we have exhibited maps in three and
four dimensions which generalise the known two-dimensional discrete
integrable maps.

The simple scheme we have described provides a plethora of interesting
cases which we plan to examine further, in particular to  characterise
higher dimensional invariant varieties completely, provide Lax pairs,
relate non-autonomous generalisations to the results
of~\cite{KaNaSa13}, and study the inflation process in more detail.
\section*{Acknowledgements} 
We gratefully acknowledge enlightening correspondence with Ivan
Cheltsov (University of Edinburgh) on geometric properties of
3-dimensional manifolds.  NJ would like to thank the LPTHE,
Universit\'e Pierre et Marie Curie, Sorbonne Universit\'e, Paris for
their hospitality while carrying out the research reported here.

\section*{Funding}
Nalini Joshi received funding through an Australian Laureate
Fellowship Grant \#FL120100094 from the Australian Research Council
while completing this work.

\providecommand{\bysame}{\leavevmode\hbox to3em{\hrulefill}\thinspace}
\providecommand{\MR}{\relax\ifhmode\unskip\space\fi MR }
\providecommand{\MRhref}[2]{%
  \href{http://www.ams.org/mathscinet-getitem?mr=#1}{#2}
}
\providecommand{\href}[2]{#2}

\end{document}